\documentclass[twocolumn,twoside,hyphens]{Jornadas}
\usepackage[utf8]{inputenc}
\usepackage[english]{babel}
\usepackage{listings}

\usepackage{algorithm}
\usepackage{algorithmicx}
\usepackage{algcompatible}
\usepackage{adjustbox}
\usepackage{graphicx}
\usepackage{color}
\usepackage{caption}
\usepackage[caption=false,font=footnotesize]{subfig}

\usepackage{placeins}
\usepackage{hyperref}
\usepackage{url}
\hypersetup{breaklinks=true}

\definecolor{gray97}{gray}{.97}
\definecolor{gray75}{gray}{.75}
\definecolor{gray45}{gray}{.45}

\def\BibTeX{{\rm B\kern-.05em{\sc i\kern-.025em b}\kern-.08em
    T\kern-.1667em\lower.7ex\hbox{E}\kern-.125emX}}

\graphicspath{{.}{./Figuras/}}

\usepackage{booktabs}
\usepackage{makecell}
\usepackage{textcomp}
\usepackage{xspace}
\newcommand{\fig}{Fig.~}
\newcommand{\taprio}[0]{{\tt taprio}\xspace}

\newcommand{\etf}{{\tt etf}\xspace}
\newcommand{\netem}{{\tt netem}\xspace}
\newcommand{\clsact}{{\tt clsact}\xspace}
\newcommand{\veth}{{\tt veth}\xspace}
\newcommand{\skb}{{\tt sk\_buff}\xspace}
\newcommand{\skbprio}{{\tt sk\_buff$\rightarrow$priority}\xspace}
\newcommand{\afp}{{\tt \footnotesize AF\_PACKET}\xspace}
\newcommand{\afi}{{\tt \footnotesize AF\_INET}\xspace}
\newcommand{\xdp}{{\tt \footnotesize XDP}\xspace}
\newcommand{\clockrt}{\texttt{\footnotesize CLOCK\_REALTIME}\xspace}
\newcommand{\clockmono}{\texttt{\footnotesize CLOCK\_MONOTONIC}\xspace}

\begin{document}

\title{Characterization of latency and jitter in TSN emulation}

\author{%
   Álex Gracia%
   \thanks{Dept. of Computing and Systems Engineering - I3A, Universidad de Zaragoza, \\e-mail: {\tt \{alex.gracia,briz,jsegarra\}@unizar.es}},
   José Luis Briz%
   \footnotemark[1],
     Héctor Blanco-Alcaine%
   \thanks{Intel Corporation | Intel Deutschland GmbH, \\e-mail: {\tt \{hector.blanco.alcaine@intel.com}},
   Juan Segarra%
   \footnotemark[1],
   Alitzel G. Torres-Macías%
   \textsuperscript{1,}\thanks{CINVESTAV Unidad Guadalajara, México, \\e-mail: {\tt \{alitzel.torres,antonio.ramirezt\}@cinvestav.mx}},
   Antonio Ramírez-Treviño%
   \footnotemark[3]
}

\maketitle

\markboth{}{}
\pagestyle{empty} 
\thispagestyle{empty}

\begin{abstract}
This research focuses on timestamping methods for profiling network traffic in software-based environments. Accurate timestamping is crucial for evaluating network performance, particularly in Time-Sensitive Networking (TSN). We explore and compare four timestamping techniques within a TSN emulation context, though its findings extend to other network scenarios. The study leverages the Mininet emulator to model TSN networks, defining hosts, bridges, links, and traffic streams. It characterizes bridge latencies and jitter, solves the TSN scheduling problem based on measured parameters, and evaluates the correctness of a deployed schedule for a use case. Key contributions include a methodology for software-based timestamping, solutions for TSN emulation challenges in Linux and Mininet, and experimental insights for optimizing TSN emulation platforms on various system configurations, with and without Intel\textsuperscript{\tiny\textregistered}'s TCC\textsuperscript{\tiny\textregistered}, either on a high-end workstation or on an industrial PC.

\end{abstract}


\section{Introduction}
\label{sec:intro}

\PARstart{T}{imestamping} frames is central to network profiling. It is primarily performed using network analyzers in physical networks. Profiling network traffic in software has become crucial in emulated and containerized environments (e.g., Docker, Kubernetes, CNI plugins), bridging and tunneling, and cloudification. It enables the use of different procedures for recording various timer values at different software layers, each with distinct overheads and trade-offs. The goal of this work is to explore and compare four different timestamping methods. We conduct our study in the context of Time-Sensitive Networking (TSN) emulation, but our methods and findings are applicable to a wide range of network emulation scenarios, including containerized networks and tunneling systems.

Time-Sensitive Networking (TSN) constitutes a set of IEEE standards at the Data Link layer aimed at achieving deterministic and ultra-fast transmissions over standard Ethernet and wireless technologies, capable of integrating different types of traffic. Current proprietary industrial networks are migrating to this new open and interoperable paradigm. TSN is at the basis of Industry 4.0, new intra-vehicle networks in automotive and aerospace industries, and the future Deterministic Internet.

TSN offers traffic shaping to maintain quality of service in networks where time-sensitive streams with varying criticalities coexist with best-effort traffic. Particularly, we focus on the IEEE~802.1Qbv shaper (TAS)~\cite{IEEE1Q-2014}. The quest for optimal routing and scheduling solutions for specific use cases still remains open, particularly when it comes to implementing a scheduling solution in actual networks. This implementation requires a verification process that often necessitates revising the schedule due to estimates provided to the theoretical scheduling problem. Utilizing a physical testbed demands significant time and resources, and simulation may not be practical due to the complexity of certain use cases. In this research, we focus on emulation, which aligns with the trend towards Software-Defined Networks (SDN) and cloudification, where physical testbeds are not available.

\begin{figure}
    \centering
    \includegraphics[width=0.8\linewidth]{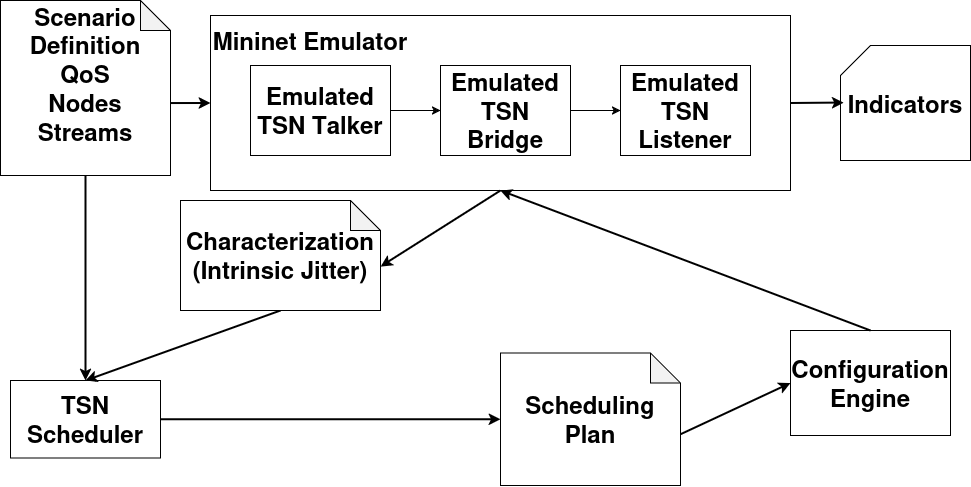}
    \caption{Overview of the experimental setup for TSN emulation using Mininet.}
    \label{fig:big-picture}
\end{figure}

To this end, we leverage the Mininet~\cite{Mininet} network emulator. It allows the definition and emulation of network nodes (\emph{hosts} ---TSN end points---, and \emph{bridges}) and links on a single Linux system. Hosts in Mininet behave as the physical ones. The net can be administered through common tools (e.g., {\tt nm}). User applications can send and receive frames through virtual interfaces such as {\tt veth}, going through Mininet bridges (which can functionally act as switches). Both the emulated and the physical system can run the same binaries. \fig\ref{fig:big-picture} provides an overview of our TSN emulation setup and highlights the issues we address here. First, we define and configure the network: hosts, bridges, links, and TSN streams (talker and listener nodes, jitter and deadline bounds). Second, we characterize bridge latencies and the intrinsic jitter of this network. Third, we solve the TSN scheduling problem for those streams, taking into account the stream definitions, and the latencies and intrinsic jitter we measured in the emulated network. Then, we deploy the schedule configuring the TAS of the bridges. Finally, we start the system, collecting indicators to assess the correctness of the schedule.

This work provides the following contributions:
\begin{itemize}
    \item A software timestamping methodology to measure end-to-end and bridge latencies, comparing options suitable for any network emulation and containerized networks, among others.
    \item Solutions to the principal issues underlying the Linux and Mininet setup for TSN emulation. As far as we know and as of this writing, this is the only work which explicitly provides pivotal configuration details for TSN emulation on Mininet.
    \item Experimental evidence leading to useful hints to conveniently setup the underlying emulation platform (Linux plus microprocessor) with and without Intel\textsuperscript{\tiny\textregistered}'s TCC\textsuperscript{\tiny\textregistered}, either on a high-end workstation or on an industrial PC.
\end{itemize}

In which follows, we review a few close contributions in Sec.~\ref{sec:related}. Secs.~\ref{sec:linux-setup} and \ref{sec:timestamping-bgd} introduce solutions to the Linux and Mininet issues related to TSN emulation, and timestamping related background. Sec.~\ref{sec:experimental.environment} describes the experimental environment. Sec.~\ref{sec:timestamping} analyzes timestamping methods to characterize latencies at different levels and jitter, considering different platform configurations. Sec.~\ref{sec:usecase} applies the characterization methodology to a Use Case, revealing and solving a few final problems. Sec.~\ref{sec:conclusions} provides conclusions and remarks.

\section{Related work}
\label{sec:related}
There is a limited number of works related to the issues of emulating TSN on Mininet~\cite{millner.2019}\cite{Model-based.wfcs.2022}. Both identify ---and do not always resolve--- integration problems of TSN components on Mininet bridges. 
The authors in~\cite{Emulation} develop a measurement methodology comparing a software implementation of the TAS with a hardware testbed, but they provide no details. 

TSN emulation on Mininet is also leveraged in~\cite{failure.handling.TSN.2021}, focusing on the SDN capabilities of Mininet rather than on the TSN mechanisms implemented in Linux. They develop a \emph{profiling} methodology, whose approach differs from that of~\cite{Emulation} but obtains similar results. Also, there is an interesting summary of most TSN utilities existing in Linux in~\cite{fejes2022tsnbuildingblockslinux}.

\begin{figure}
    \centering
    \includegraphics[width=1
    \linewidth]{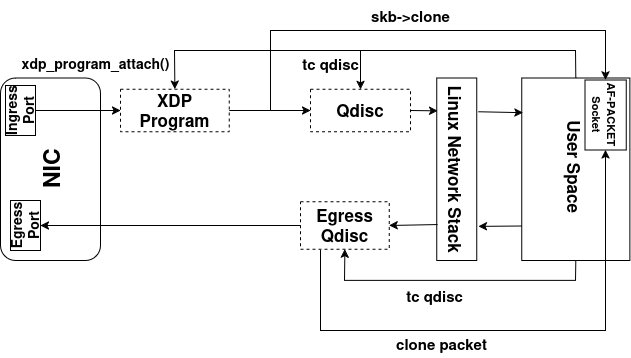}
    \caption{Flow of frames in a TSN node and the placement of the principal Linux frameworks leveraged in this research.}
    \label{fig:lns}
\end{figure}

\section{Linux Components for TSN Emulation and Timestamping}
\label{sec:linux-setup}
We exploit three key components to support TSN on Mininet: \xdp (\emph{eXpress Data Path})~\cite{xdp.seminal.2018}, \afp~\cite{afpacket} and qdisc (\emph{Linux queue disciplines}) (Fig.~\ref{fig:lns}). 

\xdp allows the execution of user-level filters using the BPF interface~\cite{BPF-Borkmann2016,BPF-LinuxKDoc}. These filters are attached to a kernel hook right after the interrupt service routine which triggers upon each frame arrival, in a physical system, or just into the \veth driver in an emulated node. The \xdp routine can examine the frame and either drop it, copy it to user space, or forward it to another Network Interface Controller (NIC). \afp clones the frame and sends the copy to a user process, while the kernel's copy of the frame (\skb) proceeds to the Linux Network Stack (LNS).

\begin{figure}
    \centering
    \includegraphics[width=0.9\linewidth]{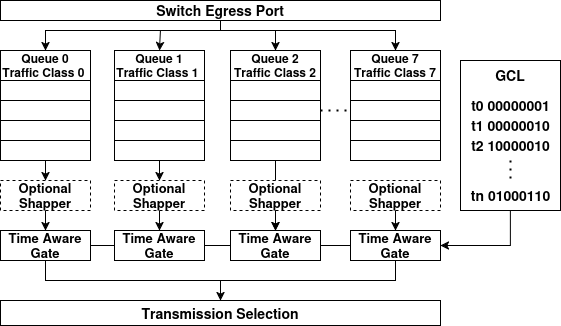}
    \caption{Structure of the IEEE 802.1Qbv Time-Aware Shaper (TAS).}
    \label{fig:TAS}
\end{figure}

Qdisc is a Linux framework, managed with the {\tt tc} tool, to place predefined filters between an ingress (egress) port and the LNS. The key qdisc for TSN is the \taprio qdisc, intended to emulate a simplified version of an IEEE~802.1Qbv TAS (\fig\ref{fig:TAS}). Besides \taprio, we also leverage \clsact for the complete integration of the \taprio qdisc, in order to meet TSN common practices (Sec.~\ref{sec:mininet-setup}), and \netem to emulate the transmission time (Sec.~\ref{sec:use-case-configuration}).

\section{Setting up Mininet for TSN}
\label{sec:mininet-setup}
Mininet runs on a single Linux computer. This obviates the emulation of IEEE~802.1AS devices to meet the TSN time-synchronization requirements, resulting in a zero clock skew. Each network node (host or bridge) is a user process, which forks children processes as required (e.g., talkers and listeners at the end-stations). Network links among nodes are set up leveraging the virtual Ethernet driver \veth. This virtual driver emulates the Data Link Layer firmware of the NICs, and serves as the OS Ethernet driver itself. All processes in a node share a single LNS and the same Linux namespace. Mininet imposes no requirement on the kernel preemption mode. We have opted for a fully preemptible kernel (RT) configuration, common in TSN nodes. 

Configuring Mininet involves installing and setting up the \taprio qdisc, which emulates the queue structure of IEEE~802.1Qbv. However, the \veth driver defaults to a single queue, which is manageable for one class but requires kernel patching to overcome this limitation~\cite{Patch-Mininet}. 
Additionally, frames in TSN need to be identified by the traffic class of their respective streams. TSN bridges commonly utilize the PCP subfield within the VLAN tag of the Ethernet frame for this purpose. Although Mininet does not natively support virtual LANs, we can modify the {\tt Host} class in the Mininet framework to enable the VLAN field.

\begin{figure}
    \centering
    \includegraphics[width=\linewidth]{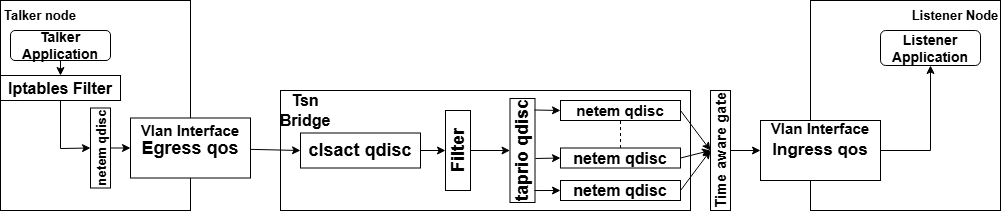}
    \caption{Linux qdiscs for TSN emulation on Mininet.}
    \label{fig:architecture.qdisc}
\end{figure}
In TSN, frames are tagged with their corresponding class ID at the talker's host. We do that leveraging the {\tt iptables} tool of Linux, which stores the ID in the priority field of the \skb allocated to the frame (\skbprio), for every stream sent through a specific egress port. Then, actual VLAN interfaces pass this ID to the PCP field of the Ethernet package. We have configured the \veth interfaces at the egress ports of Mininet hosts with VLAN support, so they can proceed the same way.

The \veth interfaces at the ingress ports of Mininet bridges cannot be configured with VLAN capabilities; otherwise, they would strip the VLAN header. Since \taprio determines the class ID of frames based on their \skbprio field, we employ the \clsact qdisc to copy the PCP value of incoming frames into their \skbprio field. \fig\ref{fig:architecture.qdisc} provides an overview of our approach, illustrating how filters and qdiscs are used to emulate TSN within Mininet.

As per time synchronization, we already mentioned that Mininet runs on a single Linux instance. Therefore, all processes can share the same clocking with no clock skew and no explicit emulation of the IEEE~802.1AS protocol.

\section{Time Coordinated Computing}
\label{sec:tcc}

Intel’s Time Coordinated Computing (TCC\textsuperscript{\tiny\textregistered})~\cite{TCC} encompasses a set of optimizations in order to improve the real-time performance of the underlying platform.
\begin{itemize}
    \item Power State Transition Optimizations limit the jitter in CPU execution due to frequency changes, and other power-saving features.
    \item Memory/Cache Allocation Optimizations reduce the variability of the memory subsystem by allowing to partition the shared caches, including the portions available to the GPU.
    \item Interrupt Request (IRQ) Optimizations streamline the critical path for interrupts in the CPU core, and also allow devices to deliver interrupts directly to the guest OS.
    \item Fabric and PCIe Virtual Channels provide different priorities for the transactions related to different workloads, allowing to treat real-time traffic as high priority.
    \item Intel\textsuperscript{\tiny\textregistered} Speed Shift for Edge Compute Applications enables specific assignment of processor performance to where it is most needed.
    \item Precision Time Coordination and PCIe Precision Time Measurement (PTM) allow to coordinate events across multiple SoC subsystems and components with independent time clocks.
\end{itemize}

\section{Timestamping Networking Events}
\label{sec:timestamping-bgd}

A usual timestamping point is the boundary between the physical layer (PHY) and the medium (e.g.,~an Ethernet cable). Specialized hardware probes like network TAP devices allow the capture of such measurements. NICs may also include timestamping mechanisms that approximate those values. Some NICs include the ability to timestamp DMA requests as well. The OS can retrieve timestamps from devices (hardware timers at the microprocessor or NICs). 

Linux stores a few hardware timestamps in the \skb per-frame structure, along with software timestamps, which can be reached leveraging \xdp, or the \texttt{recvmsg()} syscall via sockets \afp or \afi. The OS does also generate timestamps generated from OS-managed hardware timers, abstracted as OS clocks, such as \clockrt, \texttt{\footnotesize CLOCK\_TAI} or \clockmono and others, reachable through the \texttt{clock\_gettime()} syscall, or a BPF helper function such as \texttt{ bpf\_ktime\_get\_ns()} called from an \xdp program. It is pivotal to note that hardware counters may hold time values (e.g.,~in nanoseconds), or simply a number of ticks that must be translated to time values using a given frequency, as Linux does to provide clock abstractions like the ones we have just mentioned. When it comes to networking, a typical hardware element used for timestamping is the PTP Hardware Clock (PHC)~\cite{IEEE1588-2008}. Linux offers an \texttt{ioctl} interface that allows to relate the timestamps taken by the NIC and its own time-keeping mechanisms.

The hardware architecture provides specific timestamping mechanisms and ISA interfaces. For example, Intel 64 and IA-32 architectures define the operation of a \emph{Timestamp Counter} (TSC), and instructions like \texttt{rdtsc} to read it~\cite{TSC}. The OS will normally use such architecture support as the foundation of its own time-keeping mechanisms.

\section{Experimental environment}
\label{sec:experimental.environment}

Tab.~\ref{tab:experimental.hardware} summarizes the three experimental setups we use in this work. C1 employs a preeemptable kernel configuration, with no special optimization for real-time (RT), whereas C2 and C3 run a kernel with the {\tt \footnotesize PREEMPT\_RT} patch, configured with full RT preemption. The kernel in C2 is parameterized following Intel\textsuperscript{\tiny\textregistered}'s recommendations for RT. C3 runs with the Intel\textsuperscript{\tiny\textregistered}'s TCC\textsuperscript{\tiny\textregistered} system activated.

We deploy 1000 random frames from talker to listener traversing two bridges for the characterization and evaluation measurements performed in Sec.~\ref{sec:timestamping}, with \taprio configured with all queues open to avoid delaying any frame. The TSN topology and streams of the use case are described in Sec.~\ref{sec:usecase}, with \taprio configured according to the computed schedule. The default configuration is C2 unless stated otherwise.

\begin{table}[t]
\caption{Experimental platforms. Linux distribution: Ubuntu~20.04 TLS, kernel 5.2.21 in all cases.} 
\label{tab:experimental.hardware}
\begin{center}
\begin{footnotesize}

\begin{tabular}{@{}cp{31mm}cc@{}}
\toprule
Config. & \multicolumn{1}{c}{CPU} & \makecell[c]{RT\\Opt.}& {Preemption}\\
\midrule

C1 & 

Intel\textsuperscript{\tiny\textregistered} Xeon\textsuperscript{\tiny\textregistered} Gold 5120 
CPU @ 2.20GHz 56 cores 
(core \emph{Skylake}) 
& No & \makecell[c]{\emph{Preemptible}}\\

C2 & 

Intel\textsuperscript{\tiny\textregistered} Xeon\textsuperscript{\tiny\textregistered} Gold 5120 
CPU @ 2.20GHz 56 cores 
(core \emph{Skylake}) 
& Soft & \makecell[tc]{\footnotesize PREEMPT\_RT \\ \emph{full preemption}}\\

C3 & 

IEi DRPC-240 11th Gen Intel\textsuperscript{\tiny \textregistered} {Core\textsuperscript{\tiny TM}} i7-1185{\tiny GRE} CPU @ 2.80GHz 4 cores (core \emph{Tigerlake}) 
& {TCC} & \makecell[tc]{{\footnotesize PREEMPT\_RT} \\
\emph{full preemption}}\\
\bottomrule
\end{tabular}
\end{footnotesize}
\end{center}
\end{table}

\section{Timestamping}
\label{sec:timestamping}
Among the timestamping possibilities introduced in Sec.~\ref{sec:timestamping-bgd}, we have shortlisted the ones compatible with the \veth framework, which actually emulates a NIC besides acting as the NIC driver, and with the timestamping points most appropriate to obtain the value of the delays under consideration.

\subsection{Timestamping points, delays and methods}
\label{sec:timestamping-methods}

\begin{figure}
    \centering
    \includegraphics[width=1\linewidth]{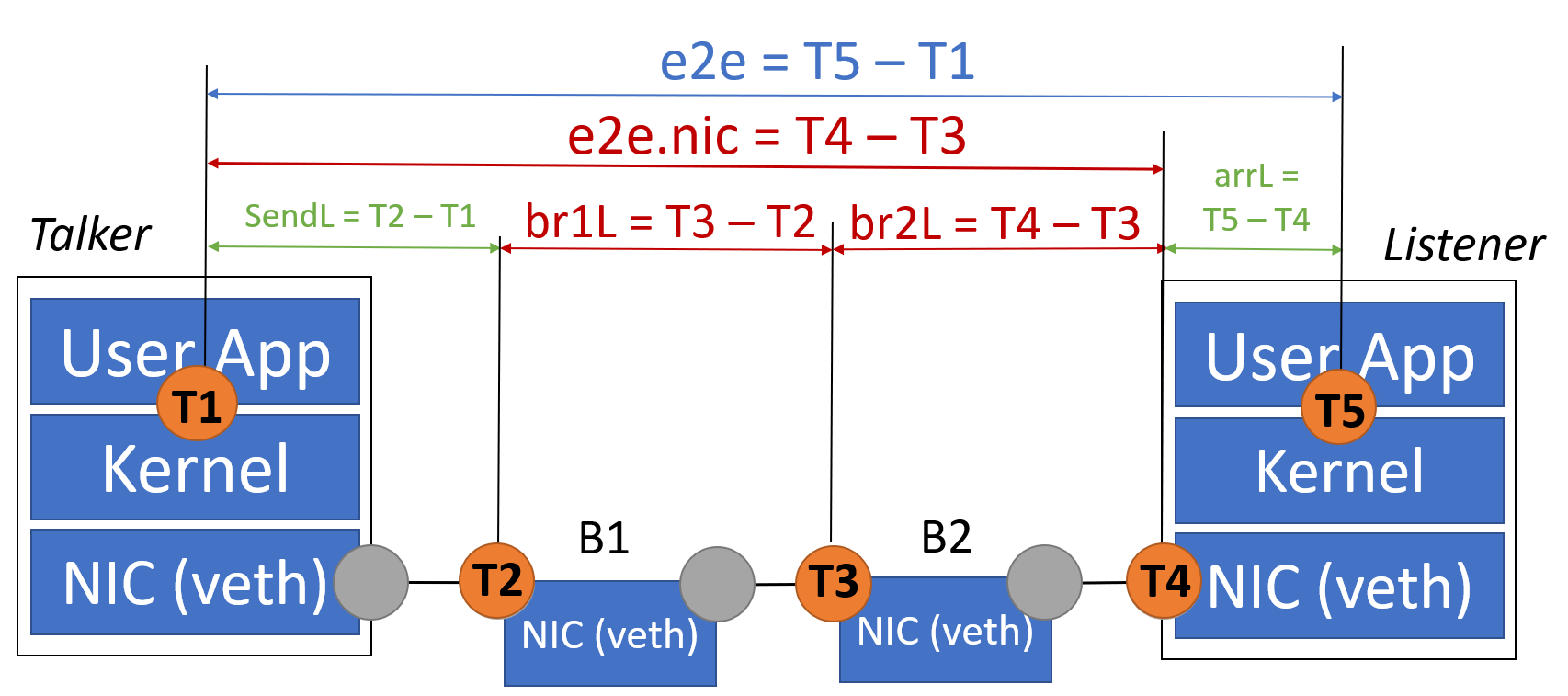}
    \caption{Timestamping points at end-stations and bridges, and latency calculation.}
    \label{fig:timestamping-points}
\end{figure}

\fig\ref{fig:timestamping-points} shows the points where we record timestamps, along with the calculated latencies. Tab.~\ref{tab:timestamping-points} defines each timestamp and the method(s) used to read them.

We leverage three timestamping methods:
\begin{description}
    \item[M1] reads the values of either {\tt \footnotesize CLOCK\_REALTIME} (M1.1) or {\tt \footnotesize CLOCK\_MONOTONIC} (M1.2) using the {\tt clock\_gettime()} syscall.

    \item[M2] records the {\tt \footnotesize CLOCK\_REALTIME} value stored in the \skb by {\tt veth\_xmit()} at \veth pairs, in kernel space.

\begin{description}
    \item [M2.1] records the timer value using a socket \afi ({\tt \footnotesize SOCK\_DGRAM}) from user space.
    \item [M2.2] records the timer value using a socket \afp from user space. \afp clones the \skb and sends a copy to user space whereas the frame proceeds as usual through the LNS, where it can be validated.
    \end{description}
    \item[M3] records the value of a {\tt \footnotesize CLOCK\_MONOTONIC} timer in kernel space, using a BPF helper function {\tt bpf\_ktime\_get\_ns()} through the \xdp framework.
\end{description}

\begin{table}[t]
    \caption{Definition and methods of the timestamps in \fig\ref{fig:timestamping-points}.}
    \label{tab:timestamping-points}
    \begin{center}
        \begin{small}
    \begin{tabular}{@{}cp{71mm}@{}}
    \toprule
    Ts & Definition and method (M) \\
    \midrule
    T1   &  Send time registered at talker's end-station M1\\
    T2 & Arrival time registered at NIC (\veth) of bridge B1. M2.2 vs. M3\\
    T3 &  Arrival time registered at NIC (\veth) of bridge B2. M2.2 vs M3\\
    T4 &  Arrival time registered at NIC (\veth) of the listener's end-station. M2.1 vs. M3\\
    T5 & Arrival time registered at the listener's end-station. M1\\
    \bottomrule
    \end{tabular}
      \end{small}
      \end{center}
\end{table}

There are two relevant latencies we must calculate for TSN scheduling. First, the definition of each time-aware stream $s_i$, in a set of $i$ time-aware streams, includes a maximum allowed delay $D_i$ and jitter $J_i$ for the stream. For all frames of stream~$s_i$, the delay $d_{i,j}$ of each frame~$j$ must be equal or lower than $D_i$. In TSN scheduling, such delay (and its jitter) refers to the time span we measure in the emulated system as \emph{e2e.nic} ($T4-T1$, \fig\ref{fig:timestamping-points}). Thus, the TSN schedule deployed in the emulated system is correct if the actual \emph{e2e.nic} value measured for every frame~$j$ is such that $d_{i,j}\leq$~\emph{e2e.nic}.

Second, in order to correctly calculate the gate opening and closing times synthesized as GCL entries in the IEEE~802.1Qbv TAS (\fig\ref{fig:TAS}), the scheduling algorithm must take into account the intrinsic jitter of the physical (or emulated) TSN system. The two main factors of this intrinsic jitter are the clock skew (which is zero in Mininet, Sec.~\ref{sec:mininet-setup}) and the bridge latency. Actual TSN bridges have no user processes. In Mininet bridges, we deploy an instrumental user process which only runs when we activate profiling, using M2.2 and M3 to measure the bridge latency (\emph{br1L} for any intermediate bridge, and \emph{br2L} for the last bridge before the listener's end station as in \fig\ref{fig:timestamping-points}).

We have also instrumented the profiling system to calculate \emph{sendL}, \emph{arrL}, and \emph{e2e}, useful to get an insight on the ways a frame can be processed in Linux. 

\subsection{AF\_PACKET (M2.2) vs. XDP (M3)}
\label{sec:afp-xdp}
\begin{figure}
    \centering
    \includegraphics[width=1\linewidth]{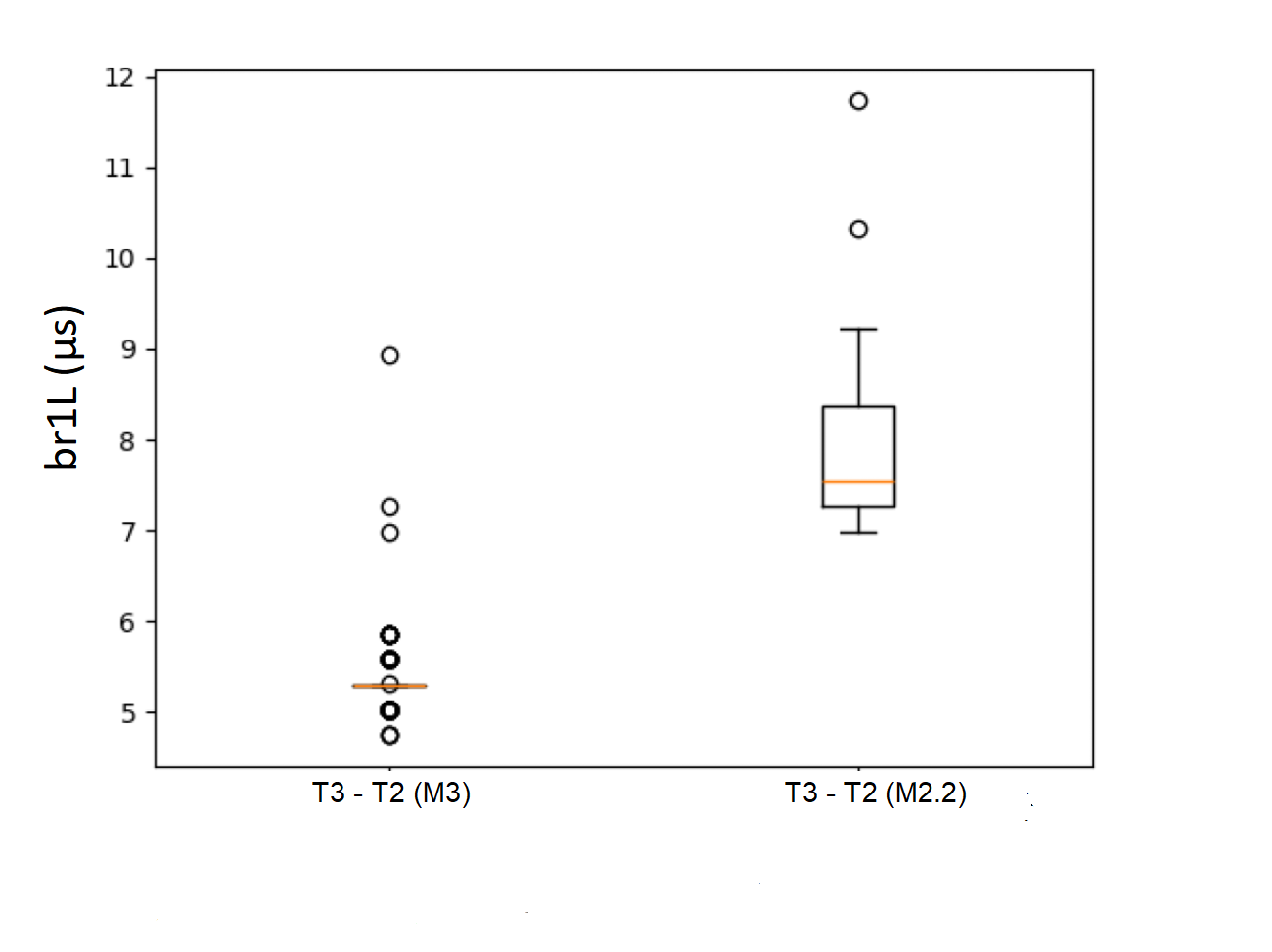}
    \caption{Bridge latency (br1L) measured with M2.2 and M3.}
    \label{fig:afp-xdp-brl}
\end{figure}

\fig\ref{fig:afp-xdp-brl} shows that the bridge latency (\emph{brL1}) and average jitter measured using M2.2 (\afp) is slightly higher and with greater IQR than using M3 (\afp), due to the cloning performed by the latter. \xdp yields a greater absolute jitter if we consider the outliers. Differences are actually negligible (a few $\mu$s), favoring \afp (M2.2) because \xdp (M3) is much harder to implement.

\begin{figure}
    \centering
    \includegraphics[width=0.9\linewidth]{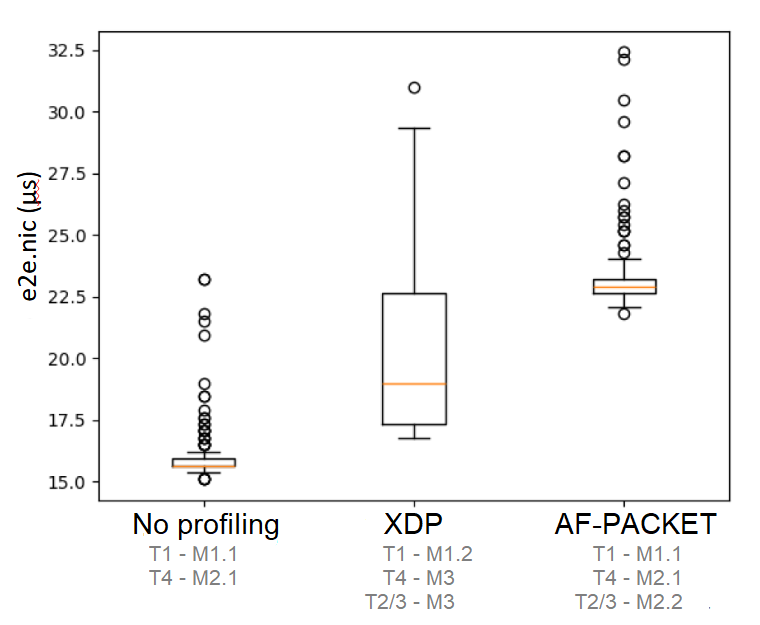}
    \caption{End-to-end latency \emph{e2e.nic} measured with M2.2 and M3.}
    \label{fig:afp-xdp-e2enic}
\end{figure}

The leftmost box plot in \fig\ref{fig:afp-xdp-e2enic} shows \emph{e2e.nic} measured using M1.1 for $T1$ and M2 for $T4$, with timestamping turned off in bridges. It provides an estimate of the overhead introduced by \xdp (M3) and \afp (M2.2) when timestamping is active at \emph{brL1} and \emph{brL2}. Again, practical differences between \xdp and \afp are negligible. Also, we can estimate the overhead of bridge timestamping while measuring \emph{end2end.nic} in about 10~$\mu$s.

\subsection{Impact of configuration}
\label{sec:impact-configuration}
\begin{figure*}
    \centering
    \includegraphics[width=1\linewidth]{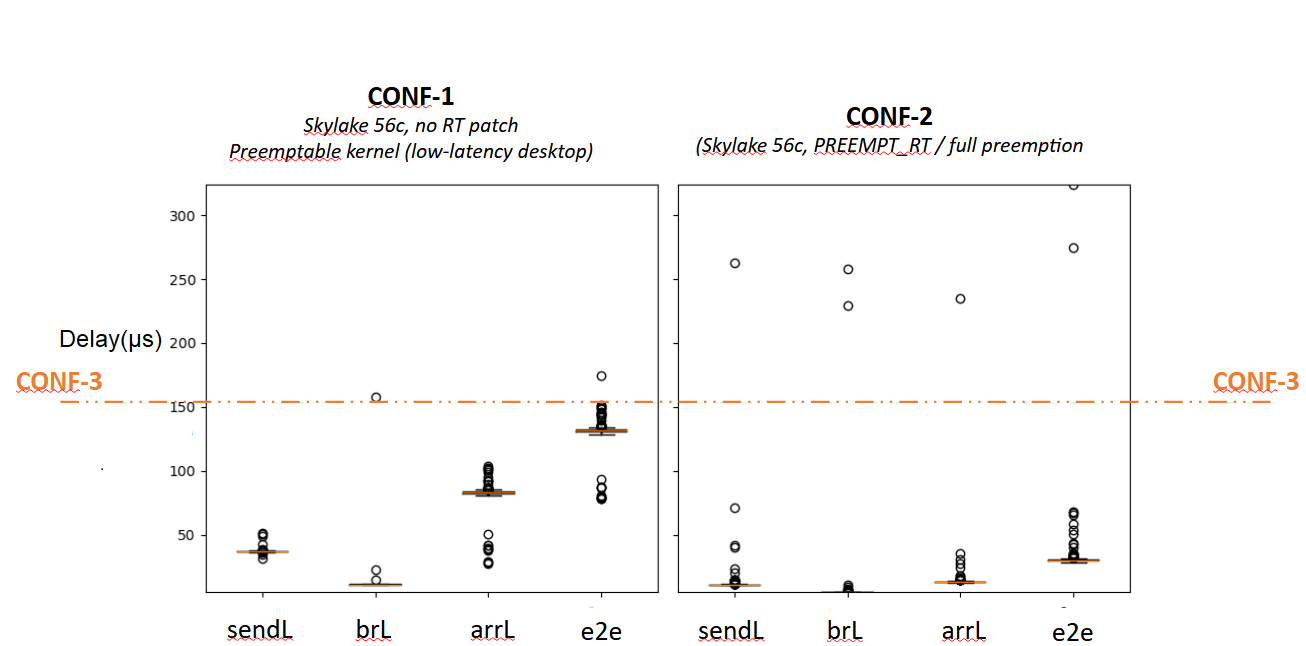}
    \caption{Impact of using configurations C1, C2, or C3 (Tab.~\ref{tab:experimental.hardware}) in latency calculations, using \xdp (M3) at $T2,T3$, and $T4$.}
    \label{fig:configuration-latencies}
\end{figure*}

\begin{figure*}
    \centering
    \includegraphics[width=0.9\linewidth]{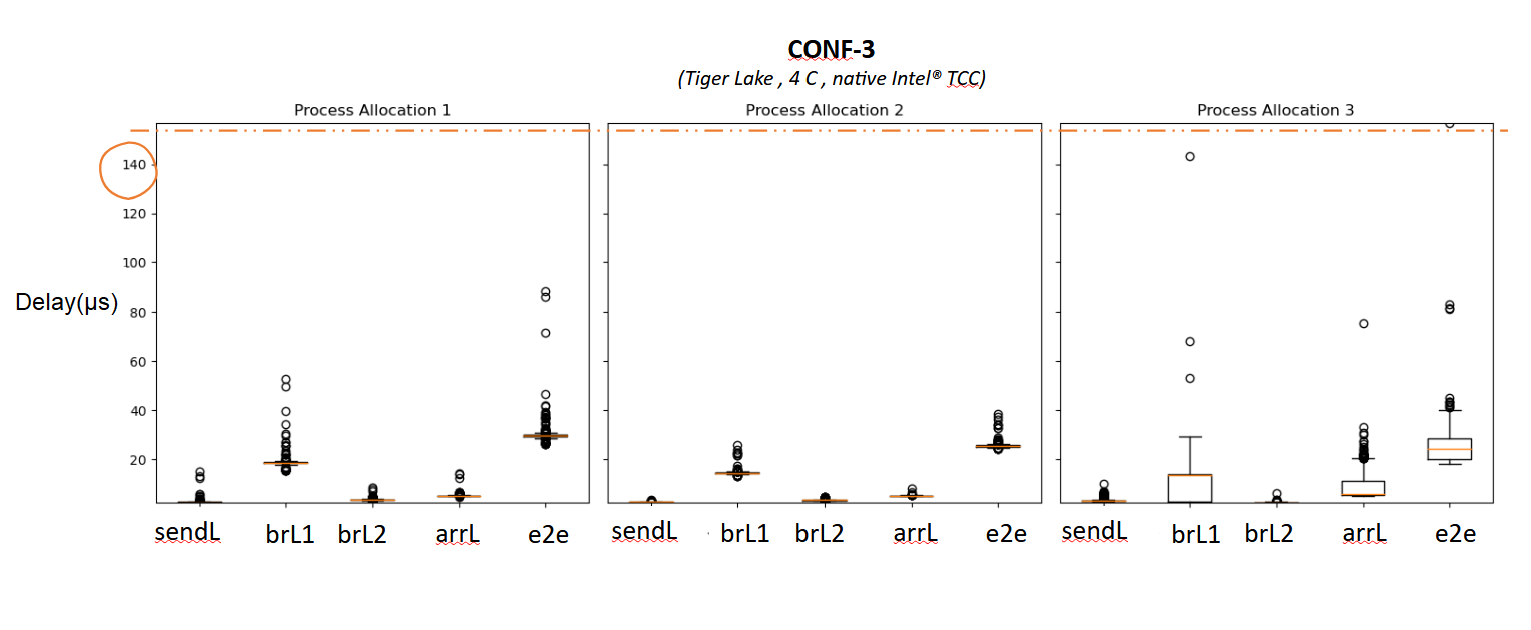}
    \caption{Impact of process allocation in latency calculations (C3, M2.2 at $T2, T3$, and $T4$).}
    \label{fig:configuration-latencies-allocation}
\end{figure*}

We have compared our methodology over the three platform configurations summarized in Tab.~\ref{tab:experimental.hardware}. The RT optimizations in C2 yield lower latency values across all timestamping methods, despite a few outliers (\fig\ref{fig:configuration-latencies}). With C3, values decrease even further, necessitating a rescaling of the $y$-axis to better visualize the impact (note the C3 dashed line at about 150~$\mu$s superimposed as a reference). We detail results with C3 in \fig\ref{fig:configuration-latencies-allocation}, marking the C3 reference given in \fig\ref{fig:configuration-latencies} (150~$\mu$s), varying now the way we allocate processes to the four cores available in C3. Although the values per latency type roughly hold no matter the allocation, the middle plot (Allocation 2) eliminates the extreme outliers seen in other allocation schemes. We use \xdp (M3) at $T2$, $T3$, and $T4$ in \fig\ref{fig:configuration-latencies} and \afp (M2.2) in \fig\ref{fig:configuration-latencies-allocation} but the results hold, with negligible variations.

\section{Use Case deployement and results}
\label{sec:usecase}
\begin{figure}
    \centering
    \includegraphics[width=1\linewidth]{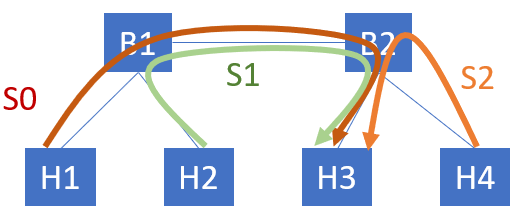}
    \caption{Use case network configuration.}
    \label{fig:use-case-network}
\end{figure}

We now deploy a simple TSN use case to achieve three key objectives (Tab.~\ref{tab:use-case-streams}, \fig\ref{fig:use-case-network}). First, we validate the timestamping methodology outlined in Sec.~\ref{sec:timestamping}. Second, we wrap-up and test the TSN emulation settings described in Secs.~\ref{sec:linux-setup} and \ref{sec:mininet-setup}. Finally, we demonstrate that the emulation platform and methodology can support the optimization of TSN scheduling design and deployment.

\begin{table}[t]
\caption{Streams of the use case under test: id, talker's and listener's hosts, Period and deadline (D)}
\label{tab:use-case-streams}
\begin{center}
\begin{tabular}{@{}ccccc@{}}
\toprule
Stream & Source & Dest. & Period & D \\
\midrule
0 & h1 & h3 & 10 ms & 10 ms \\
1 & h2 & h3 & 20 ms & 20 ms \\
2 & h4 & h3 & 30 ms & 30 ms  \\
\bottomrule
\end{tabular}
\end{center}
\end{table}

\begin{figure}
    \centering
    \includegraphics[width=1\linewidth]{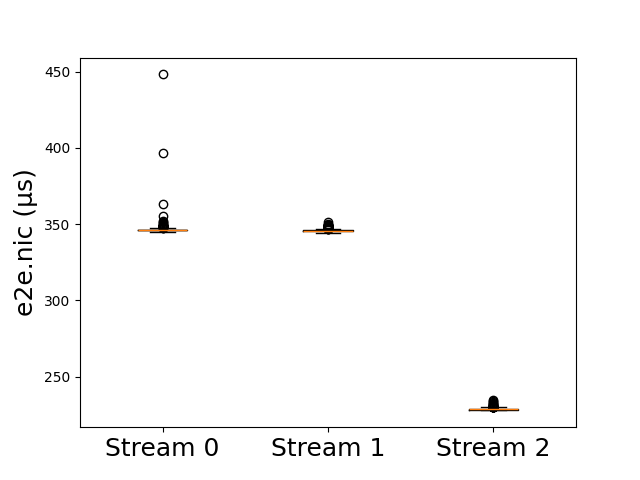}
    \caption{\emph{e2e.nic} latency for the streams in the the use case.}
    \label{fig:use-case-e2e.nic}
\end{figure}

\subsection{Stream and network scheduling parameters}
\label{sec:use-case-configuration}
We solve the TAS scheduling problem according to the method in~\cite{access.alitzel.2024}. The solver requires the four per-stream parameters shown in Tab.~\ref{tab:use-case-streams} plus their paths (traversed bridges, \fig\ref{fig:use-case-network}). 

\begin{figure*}
    \centering
    \includegraphics[width=0.6\linewidth]{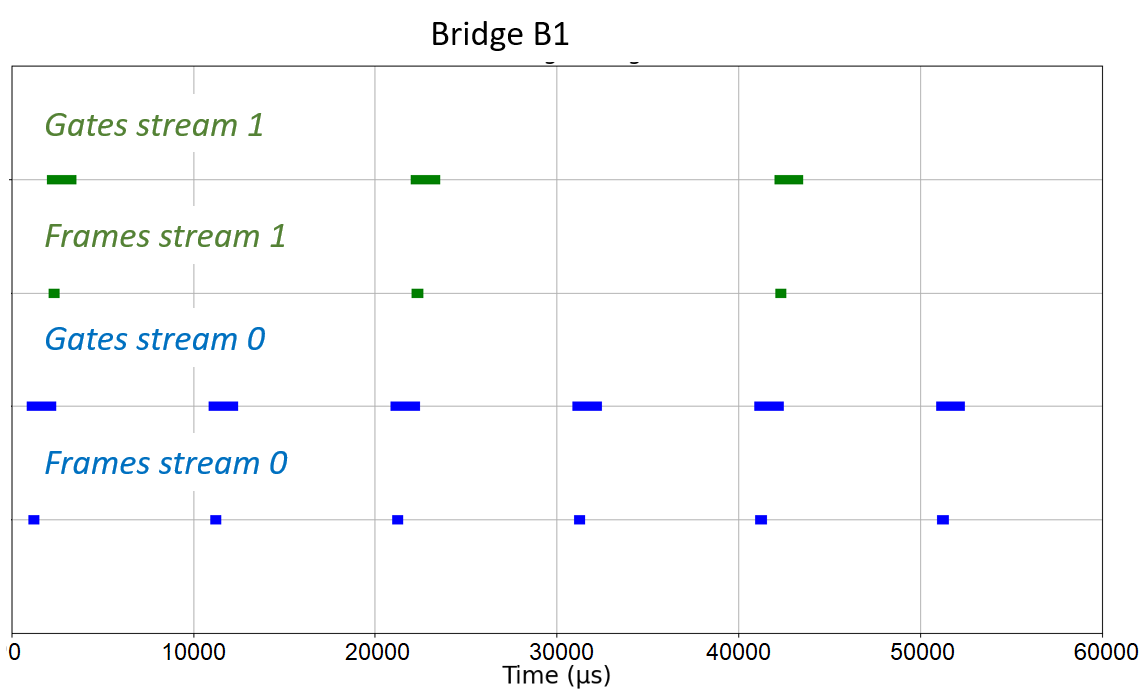}
    \caption{Pass-through times at the TAS gates of bridge B1.}
    \label{fig:pass-through-times-B1}
\end{figure*}

Also, solving the scheduling problem requires considering four network parameters: transmission time, propagation time, bridge latency, and intrinsic jitter (which includes clock skew, talker delay, and NIC/\veth jitter). The transmission time is the interval from when the Time-Aware Shaper (TAS) begins transmitting a frame through an open gate to the physical medium until the transmission completes (\fig\ref{fig:TAS}). The propagation time (also known as propagation delay) is the time it takes for a signal to travel from the sender to the receiver across a physical communication channel.

In Mininet, the \veth driver (Sec.~\ref{sec:linux-setup}) emulates the physical channel, meaning any parameter related to the latter may require kernel modifications. This is the only method we have found to accurately emulate propagation time. However, since propagation time is not central to our goals, we choose to omit it in our platform. Instead, we have devised a workaround to emulate transmission time by instantiating a \netem qdisc as a child of \taprio, as illustrated in \fig\ref{fig:architecture.qdisc}. A drawback of this approach is that the first frame entering each \netem queue is lost. Specifically, the first frame of stream~0 in \fig\ref{fig:use-case-network} disappears at B1, and the second disappears at B2. To mitigate this, we perform a dry run before starting the profiling process.

As stated in Sec.~\ref{sec:mininet-setup}, there is no clock skew in this experimental platform. To measure the talker delay, we have recorded the actual transmission times at talkers' ends-tations, obtaining that the maximum difference with the scheduled transmission times is about $80$~ns. The jitter at bridges is of about $200~\mu$s (\fig\ref{fig:afp-xdp-brl}), and the jitter in \emph{e2e.nic} latency is around $10~\mu$s. Upon these figures, we estimate in $500~\mu$s the intrinsic jitter of our Mininet emulation platform.

\subsection{System configuration and deployment}
Physical TSN systems are usually configured and run through a user network interface known as CUC (\emph{Centralized User Configuration}), operating upon a \emph{Centralized Network Conﬁguration} component (CNC) which performs the actual configuration and boots the system. We do this job through a Python script which sets up the network (Mininet hosts, bridges and links, user processes), configures the CGLs in the \taprio qdiscs of the bridges, defines a timestamp as \emph{instant zero}, and sets all processes in waiting state. When \emph{instant zero} is reached, all processes start.

\subsection{Experimental results}
\label{sec:use-case-results}
\fig\ref{fig:use-case-e2e.nic} plots the values of \emph{e2e.nic} for the three streams of the use case (configuration C2. Timestamping activated in bridges: \afp, M2.2). All measured latencies are in the order of $\mu$s, an order of magnitude below the deadlines of the streams.

We have also checked that the time windows at TAS (\taprio) gates are wide enough for the frames to pass-through. \fig\ref{fig:pass-through-times-B1} shows that the gates at the \taprio in bridge 1 allocated to streams 1 (green) and 0 (blue) open for enough time to ensure that the frames of the streams correctly pass-through. This means that the schedule solution has correctly taken into account transmission times and the intrinsic jitter of the platform.

\section{Conclusions}
\label{sec:conclusions}
We have successfully set up a Mininet/Linux environment suitable for TSN emulation, with a timestamping methodology that allows the characterization of the latencies of the emulated network (bridge latency, intrinsic jitter). As an application example, and to complete the necessary configuration steps, we have tested the schedule of a use-case on an emulated TSN network, solving the issue of emulating transmission times leveraging the \netem Linux qdisc. Emulating the propagation delay requires the modification of the kernel, nevertheless.

Using \xdp for timestamping yields slightly better latency bounds than using \afp, although differences are negligible as far as \emph{e2e.nic} (the end-to-end latency which actually counts in TSN scheduling) is concerned.

Using \xdp for timestamping offers slightly improved latency bounds over \afp, but the differences in \emph{e2e.nic}, the end-to-end latency relevant to TSN scheduling, are minimal, and AFP is significantly easier to use. Leveraging a fully preemptible kernel along with Intel\textsuperscript{\tiny\textregistered}'s TCC\textsuperscript{\tiny\textregistered} reduces substantially the intrinsic jitter in all cases. However, outcomes depend on the process-to-core allocation scheme, which is crucial for industrial PCs with few cores.

We have experimented a number of compatibility issues when installing, configuring and adapting the necessary tools and frameworks (kernel and gcc versions, \veth, qdiscs, Mininet itself among others). Improvements are in the line of updating \veth to integrate \taprio and the \etf qdisc, emulating the transmission time and updating hardware platforms.

\section*{Acknowledgments}
This work was supported by the Spanish MCIN /AEI /10.13039 /501100011033 (grant PID2022 -136454NB-C22),by Government of Aragon (research group T58\_23R) and by Instituto de Investigación en Ingeniería de Aragón (I3A, Conv. de Ayudas a Prácticas con TFG 2023)

\bibliographystyle{Jornadas}
\bibliography{biblio}

\end{document}